\documentclass[fleqn,usenatbib]{mnras}

\usepackage{newtxtext,newtxmath}
\usepackage[T1]{fontenc}

\DeclareRobustCommand{\VAN}[3]{#2}
\let\VANthebibliography\thebibliography
\def\thebibliography{\DeclareRobustCommand{\VAN}[3]{##3}\VANthebibliography}

\usepackage{graphicx, subfigure}	% Including figure files
\usepackage{amsmath}	% Advanced maths commands
\usepackage{scalerel,tikz}
\usetikzlibrary{svg.path}
\definecolor{orcidlogocol}{HTML}{A6CE39}
\tikzset{orcidlogo/.pic={
 \fill[orcidlogocol] svg{M256,128c0,70.7-57.3,128-128,128C57.3,256,0,198.7,0,128C0,57.3,57.3,0,128,0C198.7,0,256,57.3,256,128z};
 \fill[white] svg{M86.3,186.2H70.9V79.1h15.4v48.4V186.2z}
 svg{M108.9,79.1h41.6c39.6,0,57,28.3,57,53.6c0,27.5-21.5,53.6-56.8,53.6h-41.8V79.1z M124.3,172.4h24.5c34.9,0,42.9-26.5,42.9-39.7c0-21.5-13.7-39.7-43.7-39.7h-23.7V172.4z}
 svg{M88.7,56.8c0,5.5-4.5,10.1-10.1,10.1c-5.6,0-10.1-4.6-10.1-10.1c0-5.6,4.5-10.1,10.1-10.1C84.2,46.7,88.7,51.3,88.7,56.8z};
}}
\newcommand\orcidicon[1]{\href{https://orcid.org/#1}{\mbox{\scalerel*{
\begin{tikzpicture}[yscale=-1,transform shape]
\pic{orcidlogo};
\end{tikzpicture}
}{|}}}}

\title[Constraining Planets in Young Moving Groups]{Constraints on Planets in Nearby Young Moving Groups Detectable by High-Contrast Imaging and Gaia Astrometry}

\author[A.~L. Wallace et al.]{A.~L.~Wallace$^{\orcidicon{0000-0002-6591-5290 }\,1}$\thanks{E-mail: alexander.wallace@anu.edu.au}, M.~J.~Ireland$^{\orcidicon{0000-0002-6194-043X}\, 1}$\thanks{E-mail: michael.ireland@anu.edu.au}, C.~Federrath$^{\orcidicon{0000-0002-0706-2306}\,1}$
\\
$^{1}$Research School of Astronomy \& Astrophysics, Australian National University, Canberra, ACT 2611, Australia\\}

\date{Accepted XXX. Received YYY; in original form ZZZ}

\pubyear{2015}

\begin{document}
\label{firstpage}
\pagerange{\pageref{firstpage}--\pageref{lastpage}}
\maketitle

\begin{abstract}
The formation of giant planets can be studied through direct imaging by observing planets both during and after formation.  Giant planets are expected to form either by core accretion, which is typically associated with low initial entropy (cold-start models) or by gravitational instability, associated with high initial entropy of the gas (hot-start models).  Thus, constraining the initial entropy can provide insight into a planet's formation process and determines the resultant brightness evolution.  In this study, we find that, by observing planets in nearby moving groups of known age both through direct imaging and astrometry with Gaia, it will be possible to constrain the initial entropy of giant planets.  We simulate a set of planetary systems in stars in nearby moving groups identified by BANYAN $\Sigma$ and assume a model for planet distribution consistent with radial velocity detections.  We find that Gaia should be able to detect approximately 25\% of planets in nearby moving groups greater than $\sim0.3\,M_\text{J}$.  Using 5$\sigma$ contrast limits of current and future instruments, we calculate the flux uncertainty, and using models for the evolution of the planet brightness, we convert this to an initial entropy uncertainty.  We find that future instruments such as METIS on E-ELT as well as GRAVITY and VIKiNG with VLTI should be able to constrain the entropy to within 0.5\,$k_{B}$/baryon, which implies that these instruments should be able to distinguish between hot and cold-start models.
\end{abstract}


%%%%%%%%%%%%%%%%%%%%%%%%%%%%%%%%%%%%%%%%%%%%%%%%%%

%%%%%%%%%%%%%%%%% BODY OF PAPER %%%%%%%%%%%%%%%%%%

\section{Introduction}
In order to observationally constrain the formation and evolution of planetary systems, it is necessary to observe planets during or shortly after formation.  While there has been some success with discovering planets in the process of formation (e.g., the PDS 70 system; \citet{keppler2018discovery,benisty2021circumplanetary}), recent direct imaging surveys of the nearest star-forming regions, despite discoveries of new brown dwarf companions, highlighted a low frequency of wide orbit planetary mass companions \citep{kraus2011lkca,wallace2020high}.  The low number of positive detections of planets in wide orbits combined with sensitivity limits has produced upper limits on the frequency of giant planets \citep{bowler2018occurrence} and constraints on formation models \citep{nielsen2019gemini,vigan2020sphere}.  From predicted occurrence rates, it has also been determined that current instruments have insufficient sensitivity at the expected separations to detect planets around solar-type stars in nearby star-forming regions \citep{wallace2019likelihood}.  However, there has been greater success with wide-separation planets around high-mass stars in young nearby moving groups such as $\beta$-Pictoris \citep{lagrange2009probable} and 51-Eridani \citep{macintosh2015discovery}.

Nearby moving groups have been studied in detail over the years \citep{torres2008young,zuckerman2011tucana,rodriguez2013galex} and recently, precise proper motion and parallax measurements of nearby stars have allowed reasonably accurate determination of membership to these groups \citep{gagne2018banyan,schneider2019acronym}.  There are at least 27 such associations within 150\,pc with ages less than $\sim$800\,Myr.  The young ages and small distances of these systems make them ideal for young planet surveys \citep{lopez2006nearest}.

The upcoming Gaia DR3 and subsequent data releases promise high-precision mass calculations for many giant, long-period exoplanets \citep{perryman2014astrometric}. A recent study of HR 8799 has already delivered results \citep{brandt2021first} and measured the mass of HR 8799 e.  However, Gaia's expected 5--10\,year mission lifetime puts an upper limit on the semi-major axes of Gaia-detectable planets with non-degenerate solutions, which limits the possibilities for high-constrast imaging studies.  In order to conduct high-contrast imaging studies of Gaia-detectable planets, we must observe young planets in nearby systems.

Planets in the process of formation radiate with a luminosity proportional to their accretion rate and  total mass \citep{zhu2015accreting}.  The amount of energy radiated away during formation has an effect on the internal entropy of the planet.  If the accretion shock radiates all accretion energy away, the planet will form with low entropy (cold-start).  If none of the accretion energy is radiated away, the planet will have high entropy (hot-start) \citep{berardo2017evolution}. The values of internal entropy corresponding to hot and cold-starts depend on planetary mass (as shown in the `tuning fork' diagram from \citet{marley2007luminosity}.) Hot-start planets are assumed to form quickly whereas cold-start planets gradually accrete gas through the accretion shock \citep{mordasini2012formation}.  Thus, the initial entropy of a planet can indicate its formation conditions.

After formation, planets cool and fade over time, but the rate of cooling depends on their internal entropy \citep{spiegel2012spectral}.  A hot-start planet will be brighter than a cold-start planet shortly after formation.  The brightness of a planet can then be used to determine the initial entropy.  As planets age, the luminosity decreases at a rate dependant on initial entropy.  The luminosity of hot-start planets decreases faster than cold-start planets, which means, as planets age, information about the initial entropy is lost.  However, if planets of known mass are observed at young ages, it should be possible to infer the initial entropy based on the observed flux \citep{marleau2014constraining}.

If the mass and age of a planet is known with reasonable precision, the greatest uncertainty results from the observed flux measurement.  This flux uncertainty depends on the sensitivity of our instruments.  In this study, we consider current instrument such as NIRC2, NaCo, SPHERE and GRAVITY and future instruments such as JWST, VIKiNG (interferometric instrument using VLTI), MICADO on the VLT and METIS on the E-ELT.  These instruments have observed and theoretical detection limits which we convert to a flux uncertainty.  Using models linking flux, mass and age to initial entropy, we convert this to an entropy uncertainty to determine how well the initial entropy can be constrained.

The rest of the paper is organised as follows. Section~\ref{sec:samp} summarises our stellar and planet sample and models for the evolution of planet luminosity and its dependence on entropy.  Section~\ref{sec:detect} focuses on detection limits of astrometry and direct imaging and Section~\ref{sec:simulation} presents the numbers of detectable planets in our sample by both methods.  Section~\ref{sec:constrain} explains how we can constrain the initial entropy of planets if we know the mass, magnitude and age.  Our conclusions are presented in Section~\ref{sec:conclusion}.

% ===================================
\section{Stellar and Planetary Properties}
\label{sec:samp}
\subsection{Stellar Sample}
\label{sec:star_samp}
Our stellar sample comes from nearby young (<800\,Myr) moving groups which are promising targets for planet surveys.  The stars are initially selected from  Gaia's second data release (DR2) \citep{brown2018gaia} and then sorted into moving groups using the BANYAN\,$\Sigma$ from \citet{gagne2018banyan}.  Our initial sample from Gaia DR2 included stars across the entire sky within 70\,pc, brighter than a G-magnitude of 13 and temperature greater than 3800\,K.  Beyond $\sim$70\,pc the majority of giant planets in $\sim$10\,year orbits are not detectable and moving group knowledge is less complete.  Stars cooler than 3800\,K are no-longer considered solar-type in this paper as they have a measured smaller fraction of planets \citep{johnson2007new}.  The magnitude cutoff excludes stars considered too faint for reliable adaptive optics.

\begin{figure*}
    \centering
    \includegraphics[width=1.0\linewidth]{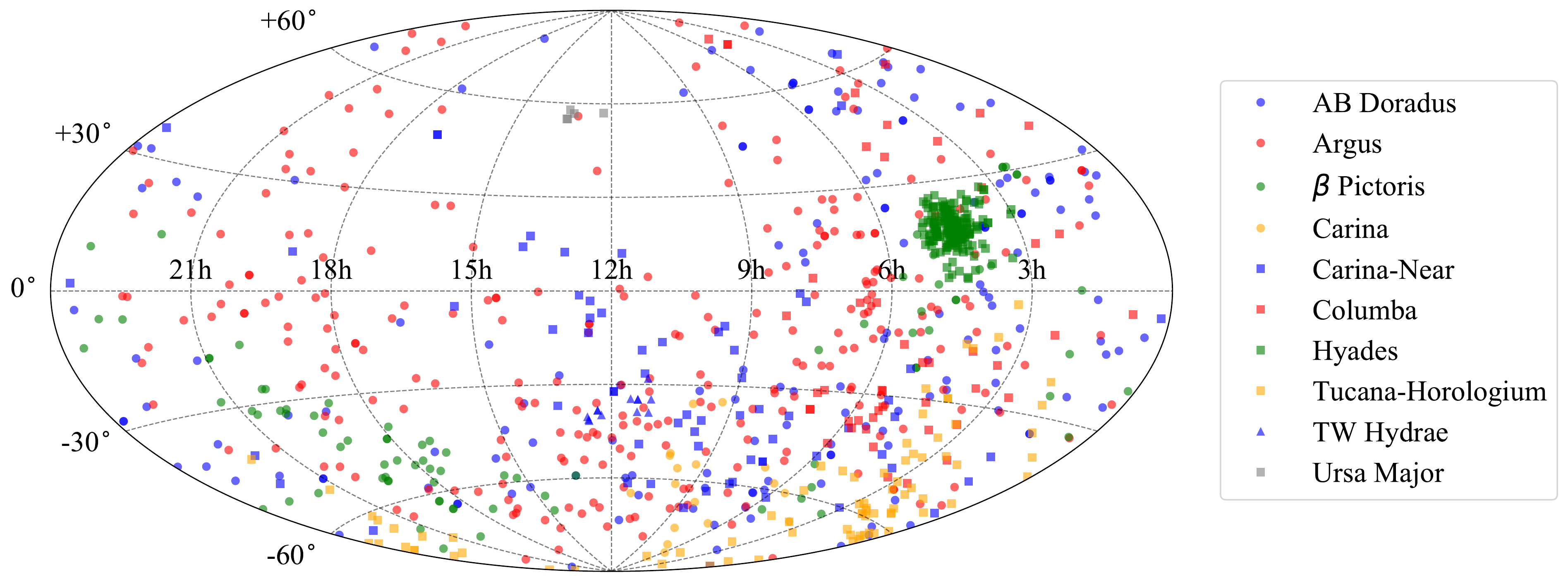}
    \caption{Stars in our sample belonging to moving groups determined by BANYAN\,$\Sigma$.  Map is shown in celestial coordinates with East to the left.}
    \label{fig:allSky}
\end{figure*}

The moving group membership of each star is then determined by the BANYAN\,$\Sigma$ algorithm.  This process uses each star's celestial coordinates, distance and proper motion values and associated errors to determine the probability of membership to a particular moving group from a list of 27 possible groups.  A star that does not belong to any group with more than 70\,\% probability is discarded from the sample.  A partial sky map of our stellar sample, indicating group membership is shown in Figure~\ref{fig:allSky}.  This map only includes targets with a membership probability greater than 95\%.

Although the BANYAN\,$\Sigma$ algorithm can sort stars into 27 moving groups, there are only 10 groups with members closer than 70\,pc as shown in Figure~\ref{fig:allSky}.  Our target stars are spread across the sky, but the majority is in the southern hemisphere.  The Argus association has the highest number of targets, but the existence of this group has been controversial \citep{bell2015self}, as it was unclear whether it represented a single moving group.  However, recent studies have suggested the association does indeed exist with an age of 40--50\,Myr \citep{zuckerman2018nearby} so we include it in this study.

\begin{figure}
    \centering
     \subfigure[Distribution of Stellar Distance]{\label{fig:star_dist} \includegraphics[width=0.9\linewidth]{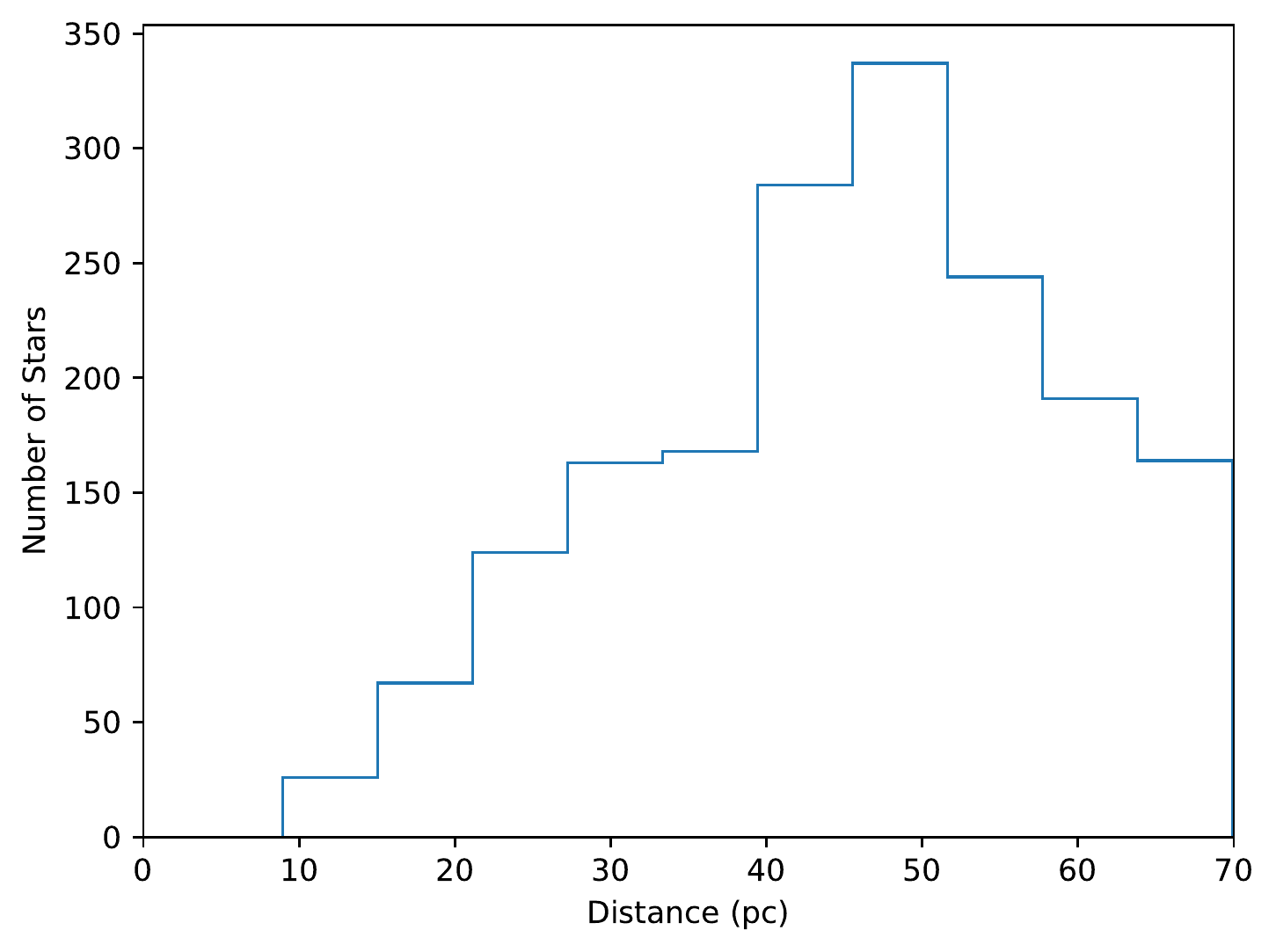}}
    \subfigure[Distribution of Stellar Mass]{\label{fig:star_mass} \includegraphics[width=0.9\linewidth]{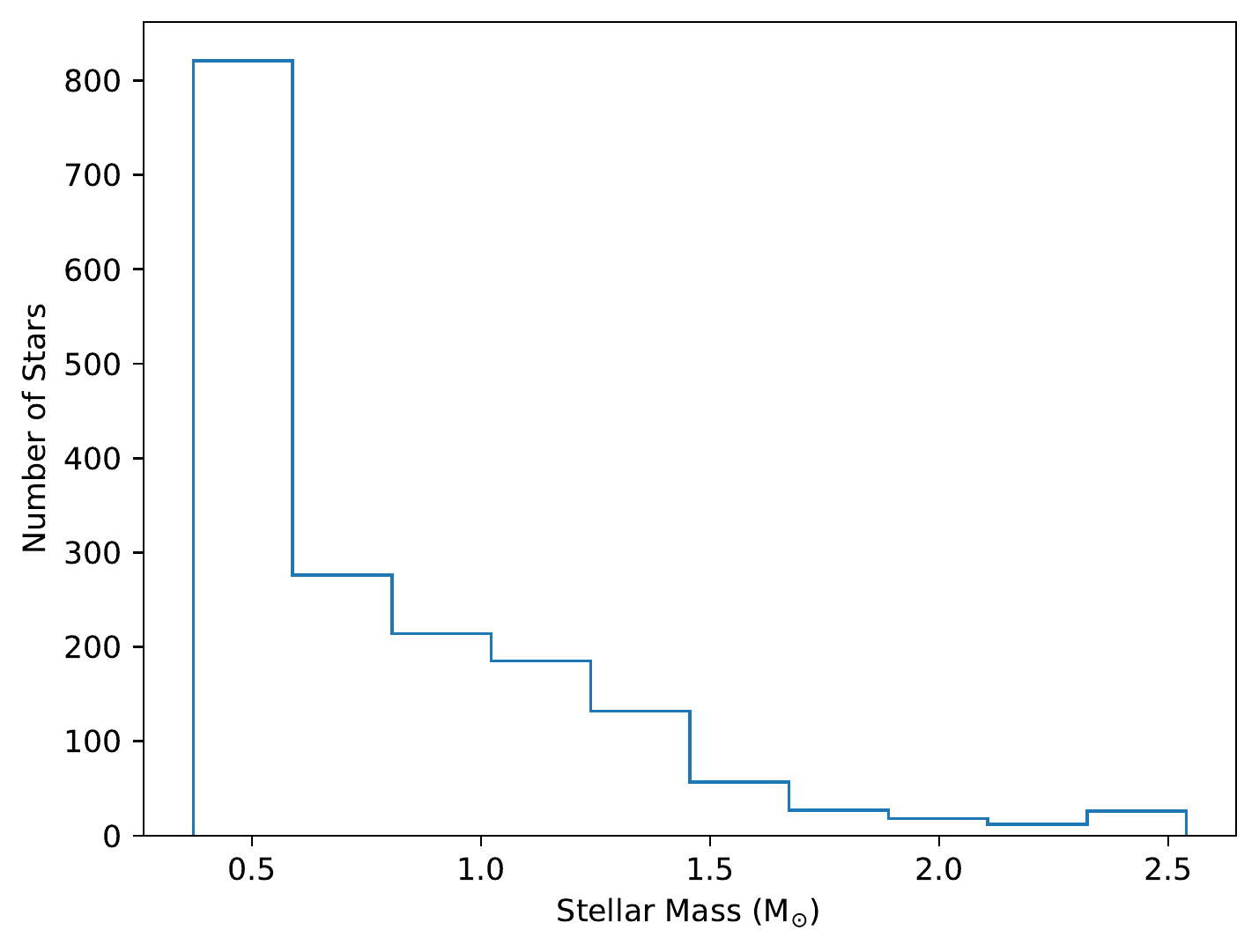}}
    \caption{Distribution of star distance and mass in our sample.  Note the distance cuts off at 70\,pc as we excluded stars further away than this.}
    \label{fig:stellar_dists}
\end{figure}

\begin{figure}
    \centering
    \includegraphics[width=0.9\linewidth]{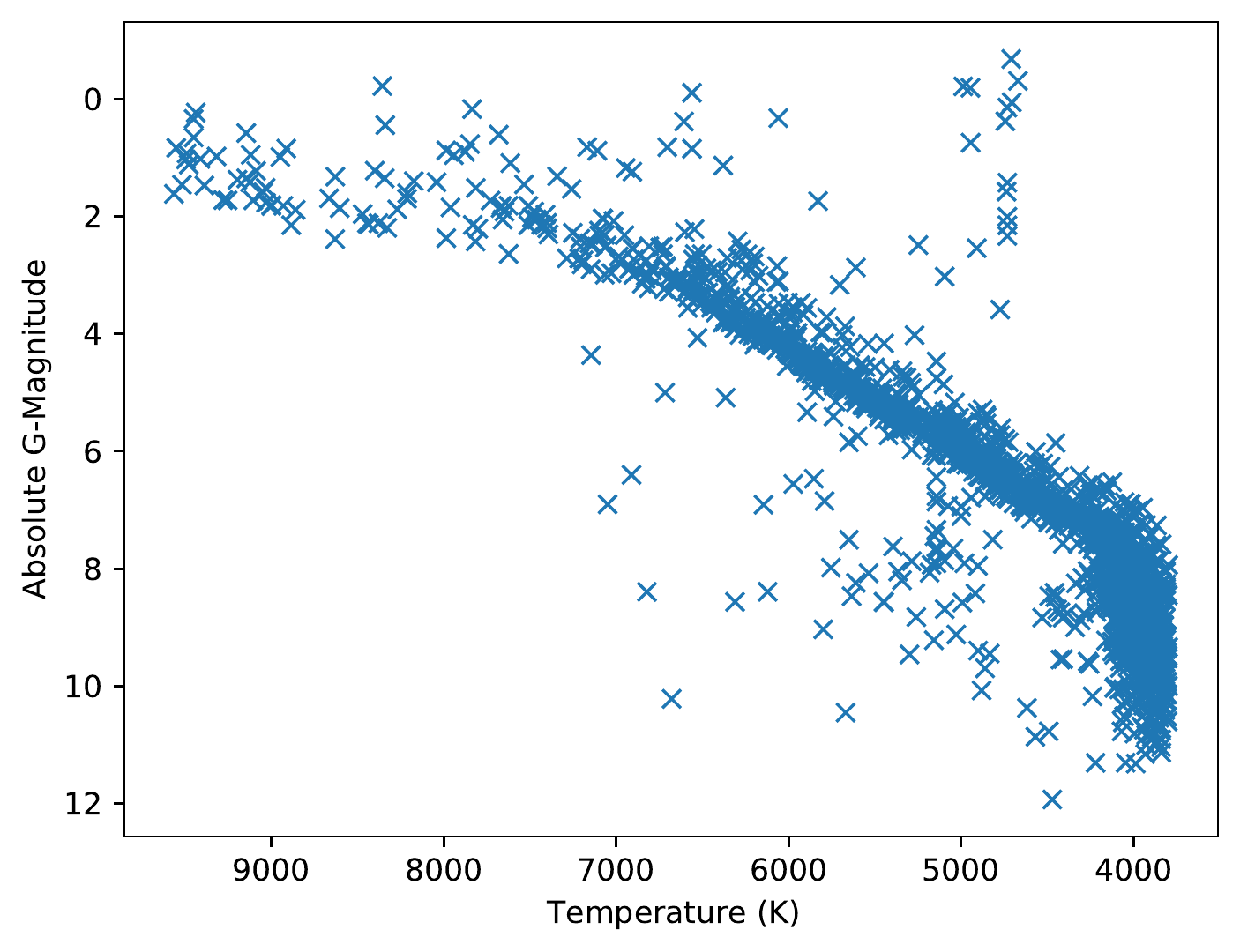}
    \caption{Colour-magnitude diagram of our targets.  The temperature scale cuts off at 3800\,K as we don't consider stars cooler than this.}
    \label{fig:hr_diagram}
\end{figure}

Our resultant sample contains 1760 stars across 10 moving groups.  The distributions of distance and mass of the stars in our sample is shown in Figure~\ref{fig:stellar_dists}.
Most of our targets are at distances of $\sim$40--60\,pc and low mass (<0.6\,M$_{\odot}$).  A colour-magnitude diagram is shown in Figure~\ref{fig:hr_diagram} which plots absolute G-magnitude against effective temperature.
As shown in Figures~\ref{fig:stellar_dists} and~\ref{fig:hr_diagram}, the majority of our targets are cool, low-mass stars which, in principle should make it easier to detect planets by astrometry.
\subsection{Planet Distribution}
\label{sec:planet_dist}
For each star in our sample, we simulate a system of planets with properties sampled from a broken power law in mass $M$ and period $P$ taken from \citet{fernandes2019hints}, which has the functional form
\begin{equation}
    \frac{d^{2}N}{d\mathrm{ln}M\,d\mathrm{ln}{P}} = CM^{\alpha}P^{\beta},
    \label{eq:power_law}
\end{equation}
where $N$ is the number of planets per star and $C$ is a normalisation constant.  We assume the mass power law is approximately consistent across all masses with $\alpha=-0.45$.  Based on radial velocity detections, the period power law changes with distance from the star with $\beta>0$ at short periods and $\beta<0$ for long periods.  This broken power law is also consistent with theoretical core formation expectations \citep{mordasini2009extrasolar}.  The study presented in \citet{fernandes2019hints} gives several different values, but in this study we use the symmetric distribution in which $\beta=0.63$ for periods less than 860\,days and $\beta=-0.63$ for periods greater than 860\,days.  The constant $C$ is set such that the total number of planets is consistent with observations.  In the symmetric distribution from \citet{fernandes2019hints},  assuming a Sun-like star, $C=0.067$.  We also note that this distribution is consistent out to 4\,au with the models based on the California Legacy Survey \citep{Fulton21}.

When simulating our planet samples, we apply the distribution from Equation~\ref{eq:power_law} in terms of semi-major axis $a$.  This changes the power-law index to 0.945 at small separations and -0.945 at wide separations (multiplying by a factor of 3/2.)  The power law changes at a fixed period of 860\,days, which corresponds to a semi-major axis of 1.77\,au for a 1\,M$_{\odot}$ star and scales with $M_{\star}^{1/3}$.  This distribution implies the majority of planets are low mass and at small separations.  While there have been high-mass planets observed at wide separations around HR~8799, $\beta$-Pic and 51~Eri, these are around high-mass stars where there is known to be an excess of high-mass planets \citep{johnson2010giant}.  The total number of planets is assumed to increase with stellar mass and some studies (e.g. \citet{bowler2009retired}) suggest a steep relationship.  However, for simplicity, we assume the number of planets scales linearly with stellar mass, as implied by \citet{Mulders2018}, and the normalisation constant $C$ is simply multiplied by $M_{\star}/M_{\odot}$.

\begin{figure}
    \centering
    \includegraphics[width=0.9\linewidth]{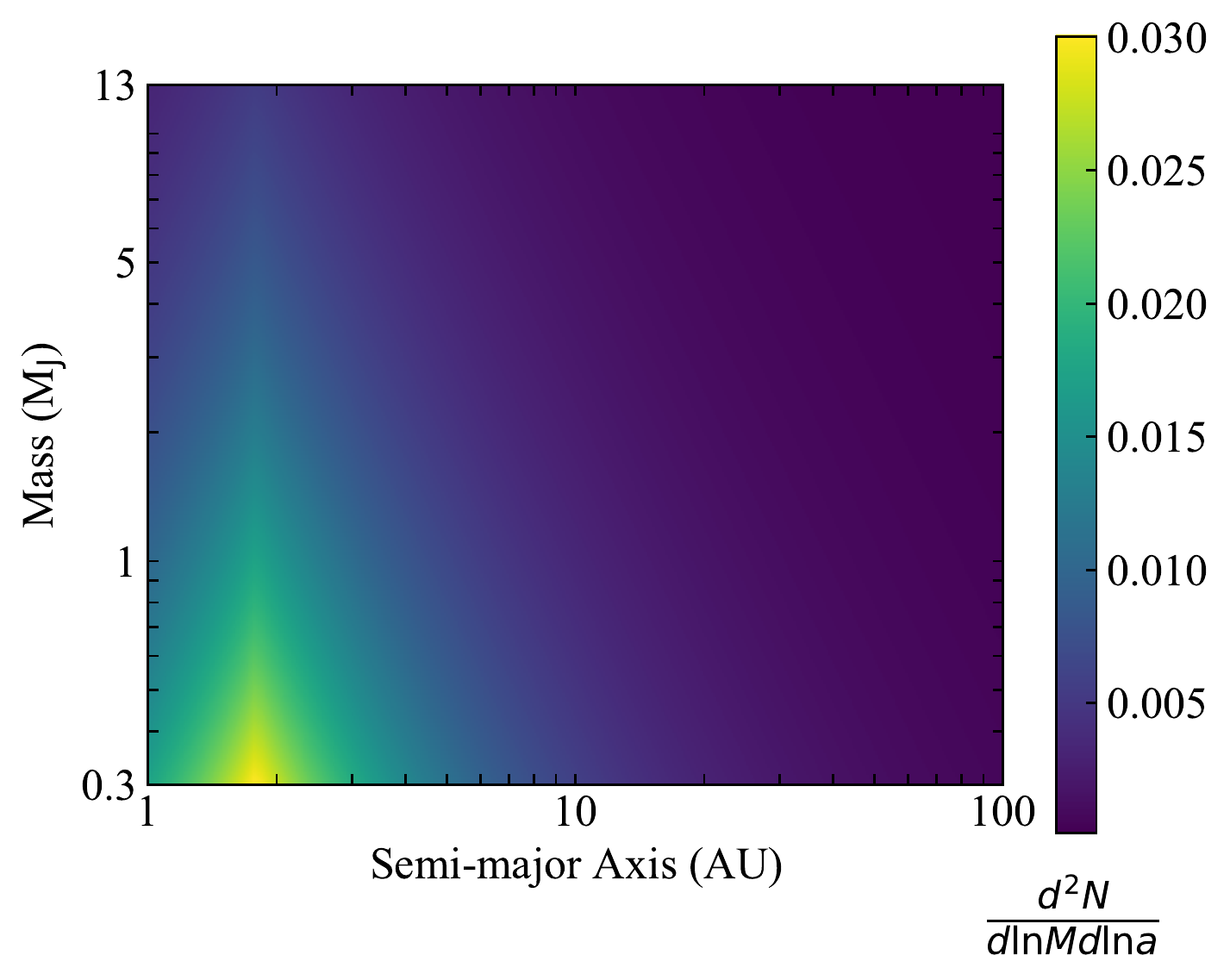}
    \caption{The differential mass and semi-major axis distribution of our simulated planets from \citet{fernandes2019hints}.  Integrating this across our simulation range gives \mbox{$N=0.08$} planets per star.}
    \label{fig:planet_dist}
\end{figure}

We simulate planet masses over a range of 0.3--13\,M$_{\rm{J}}$ and semi-major axes over a range of 1--10\,au.  Integrating the power law shown in Equation~\ref{eq:power_law} over this range gives $N\sim0.07$ planets per star.  The symmetric planet distribution from \citet{fernandes2019hints} is shown in Figure~\ref{fig:planet_dist}.  Period is converted to semi-major axis by assuming a 1\,M$_{\odot}$ star.

\subsection{Planet Luminosity and Magnitude}
\label{sec:planet_mag}
As giant planets accrete gas, the accretion energy is emitted as radiation with a luminosity proportional to the accretion rate \citep{zhu2015accreting} and are at their brightest during the period of runaway accretion \citep{lissauer2009models}.  After a planet has formed, the luminosity declines over time and is dependant on the planet's mass and entropy \citep{spiegel2012spectral}.

\begin{figure}
    \centering
    \includegraphics[width=1.0\linewidth]{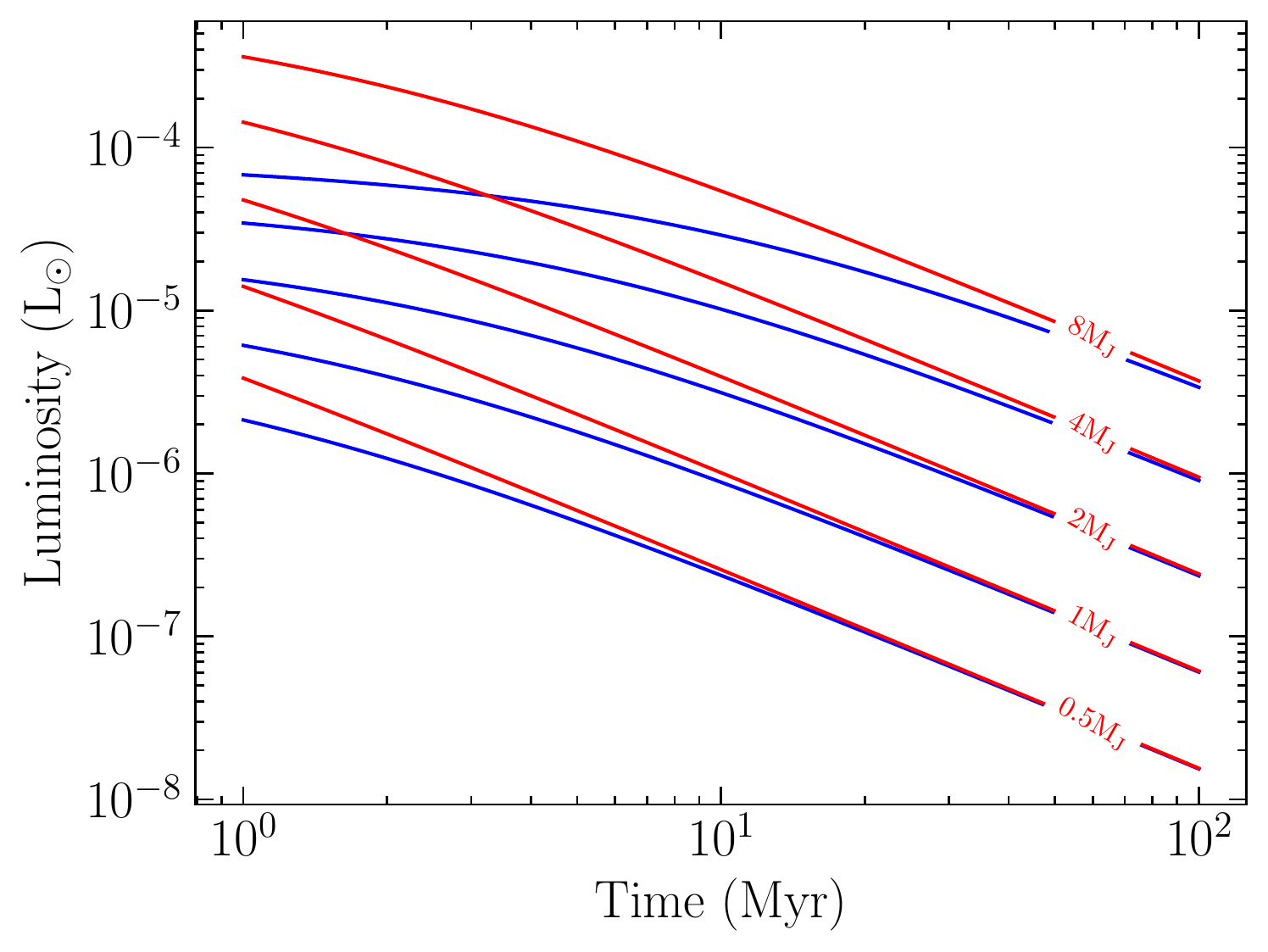}
    \caption{Evolution of planet luminosity for different planet masses and initial entropies.  The blue curves represent initial entropy of 9.5\,$k_{B}$/baryon and red represents 10.5\,$k_{B}$/baryon.  This is calculated using the temperature and radius curves from Figure~5 in \citet{spiegel2012spectral}, applying the Stefan-Boltzmann Law and interpolating between entropies shown in their paper.}
    \label{fig:lum_evol}
\end{figure}

\begin{figure}
    \centering
     \subfigure[Absolute Magnitude in K Band]{\label{fig:kmag_evol} \includegraphics[width=0.9\linewidth]{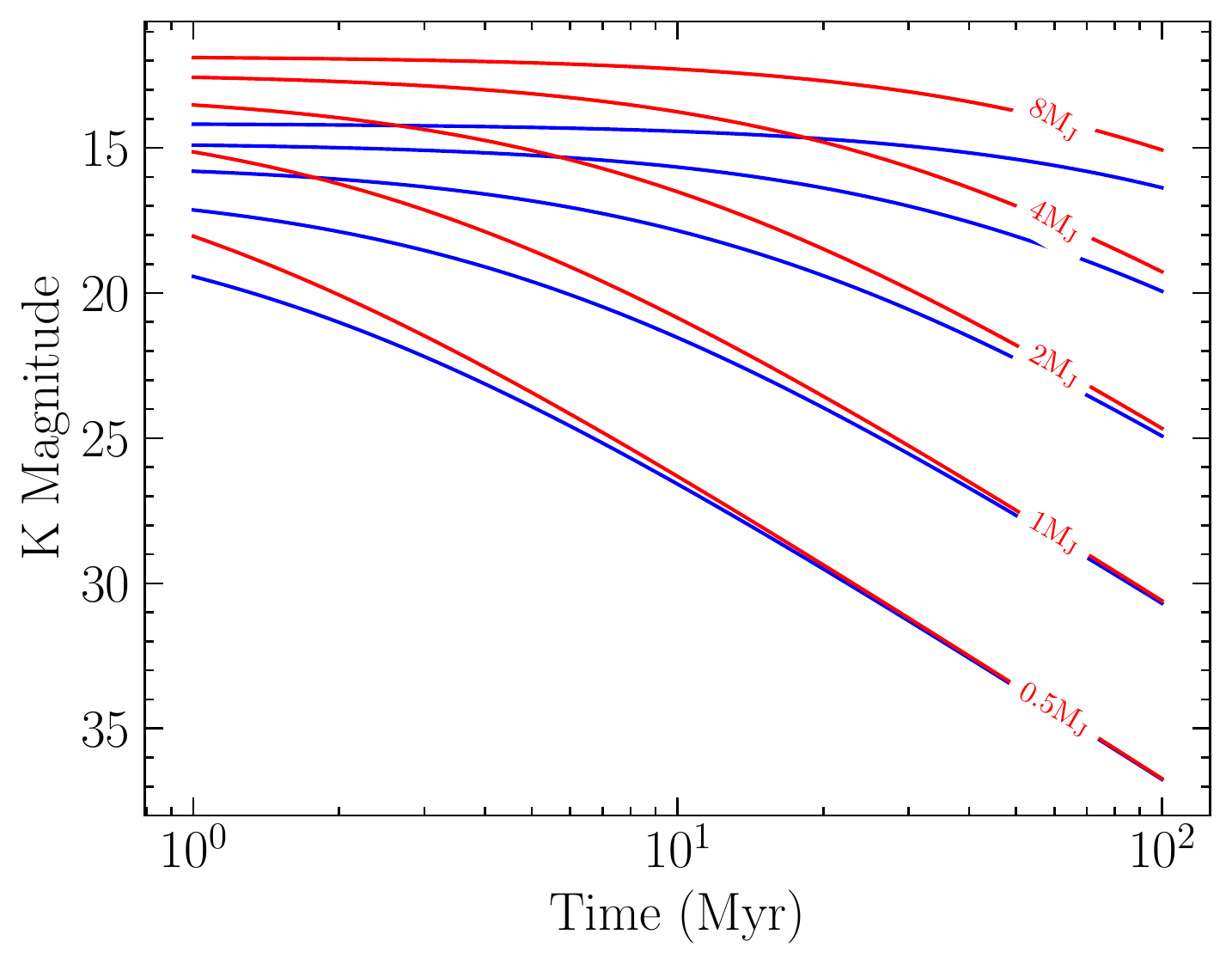}}
    \subfigure[Absolute Magnitude in L' Band]{\label{fig:lmag_evol} \includegraphics[width=0.9\linewidth]{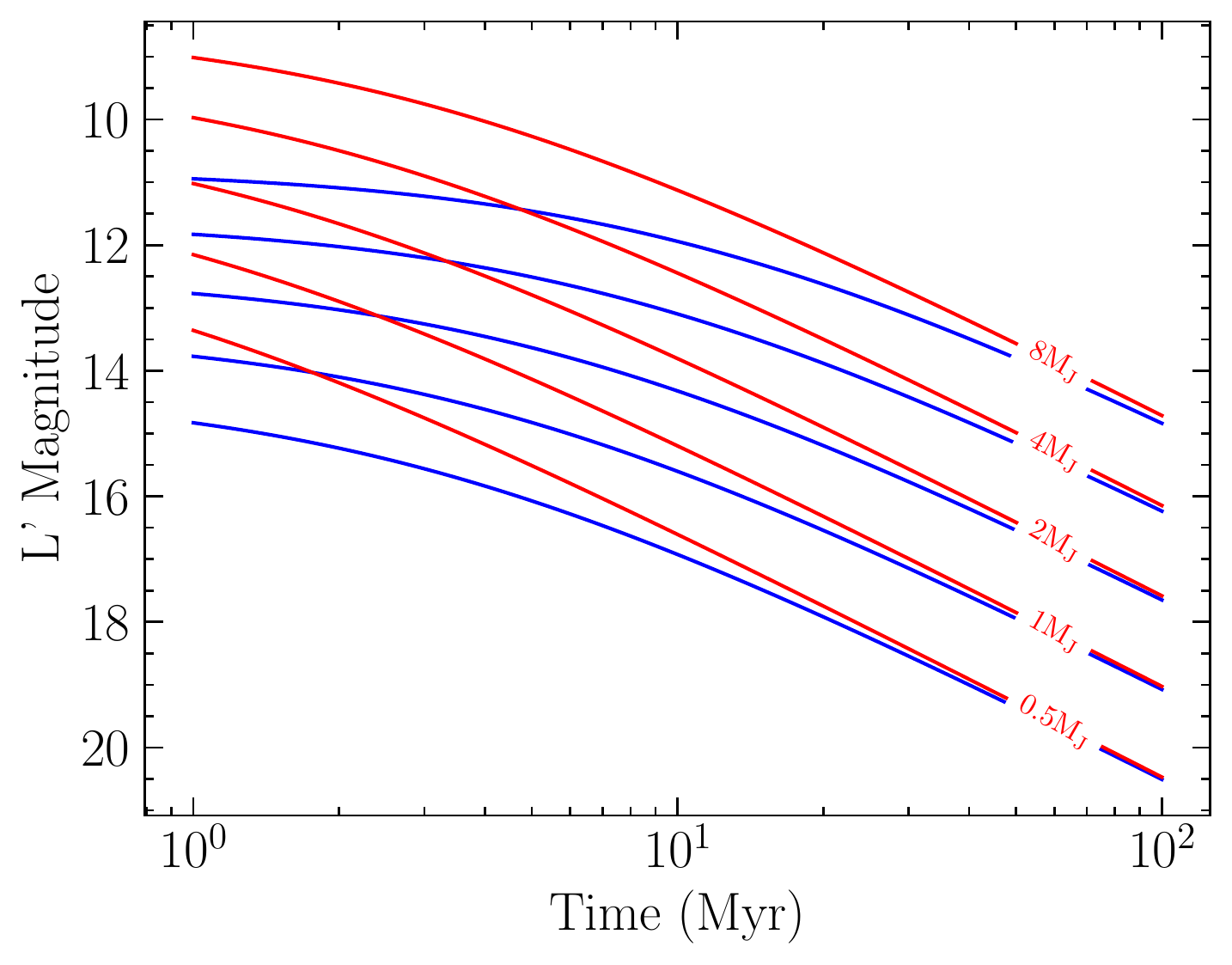}}
    \caption{Evolution of planet magnitude in the K band (panel a) and in the L' band (panel b) for different planet masses and initial entropies.  The blue curves represent initial entropy of 9.5\,$k_{\rm B}$/baryon and red represents 10.5\,$k_{\rm B}$/baryon.  This is based on the curves in Figure~7 of \citet{spiegel2012spectral} and, as before, we interpolate between entropies.}
    \label{fig:mag_evol}
\end{figure}

The conditions of planet formation have an effect on the post-formation entropy and luminosity.  If the accretion rate is slow, the planet has more time to radiate energy away, resulting in a lower internal entropy (cold-starts). Planets with low internal entropy are typically smaller and cooler and thus have lower post-formation luminosity than planets formed by hot-starts.  The luminosity as a function of age is shown in Figure~\ref{fig:lum_evol} for planets of varying mass and entropy of 9.5\,$k_{B}$/baryon (blue curves) and 10.5\,$k_{B}$/baryon (red curves) using hot and cold-start models from \citet{spiegel2012spectral}.  This is calculated from applying the Stefan-Boltzmann Law to the radius and temperature plots in Figure~5 of their paper and interpolating between their initial entropies.
The absolute magnitude evolution in near-infrared bands is also calculated in order to determine planet detectability.  This is also calculated using models from \citet{spiegel2012spectral}, taking the magnitudes from their Figure~7 and interpolating between initial entropies.  Some example magnitude evolution curves are shown in Figure~\ref{fig:mag_evol} for the K (2.2$\mu$m) and L' (3.77$\mu$m) bands using the same initial entropies as Figure~\ref{fig:lum_evol}.
The cooling curves in Figures~\ref{fig:lum_evol} and~\ref{fig:mag_evol} demonstrate that hot-start planets are initially brighter, but the luminosity declines faster.  This implies that, at old ages, hot and cold-start planets become less distinguishable as the luminosities become approximately equal.  However, at the young ages ($\sim$30\,Myr) we consider in this study, there is a noticeable difference between the two entropies for high-mass planets, which indicates we should be able to constrain the formation models of massive ($>2\,\mathrm{M_J}$) planets.

%==============================
\section{Planet Detectability}
\label{sec:detect}
\subsection{Detection by Gaia Astrometry}
\label{sec:gaia_detect}
The upcoming Gaia data releases are expected to have improved astrometry measurements of stars and the potential to discover exoplanets through this method.  
As a planet with $M_p \ll M_\star$ and its host star of mass $M_\star$ orbit their common centre of mass, the astrometric semi-major axis, also known as the astrometric signature, is given by:

\begin{equation}
    \alpha = \left(\frac{M_{p}}{M_{\star}}\right)\left(\frac{a_{p}}{d_{\star}}\right),
    \label{eq:alpha}
\end{equation}
where $a_{p}$ is the semi-major axis of the planet in au and $d_{\star}$ is the distance to the star in pc.  The planet detection capability depends on the signal to noise ratio, given by
\begin{equation}
    S/N = \frac{\alpha}{\sigma_{\rm{fov}}},
\end{equation}
where $\sigma_{\rm{fov}}$ is the accuracy per field of view crossing.  The study in \citet{perryman2014astrometric} concluded that a detection threshold of $S/N>2$ provides a reasonable estimate of planet detection numbers.  The study in \citet{ranalli2018astrometry} simulated the performance of Gaia and determined there is a 50\% chance of detecting planets in the 5 and 10\,year missions with $S/N>2.3$ and $1.7$ respectively.  For this study, we assume a planet is detectable if $S/N>2$.  The value of $\sigma_{\rm{fov}}$ depends on the star's G magnitude, but is approximately constant at 34.2\,$\mu$as for stars brighter than magnitude 12.  In this study, we use the values from Table~2 in \citet{perryman2014astrometric} and assume planets are detectable if $\alpha>2\sigma_{\rm{fov}}$.  Gaia's nominal mission is for 5\,years with a possible extension to 10\,years and this also places a constraint on detection capabilities. It is assumed that planets with periods greater than $\sim$10\,years will be poorly constrained.

\subsection{Detection by Direct Imaging}
\label{sec:direct_detect}
Direct imaging has had a lower yield for planet detection than other techniques due to the high contrast ratios between stars and planets.  Unlike transit and radial velocity, this method is more sensitive to planets at wide separations, which are less abundant.  Planets at wide separations are typically found using a combination of angular differential imaging (ADI) and reference star differential imaging (RDI) \citep{marois2006angular,lafreniere2007new} to remove instrumental artefacts combined with coronagraphy to block out the light of the central star.

Direct imaging methods are less sensitive at small separations, but this is gradually improving with new analysis methods such as kernel phase \citep{martinache2010kernel} and observational methods such as interferometry \citep{lacour2019first}.  The detection capability of an instrument is limited by both the resolution and the maximum signal to noise ratio.  This provides a `contrast limit', which typically improves with distance from the star.

\begin{figure}
    \centering
    \subfigure[K band]{\label{fig:k_limits}\includegraphics[width=0.9\linewidth]{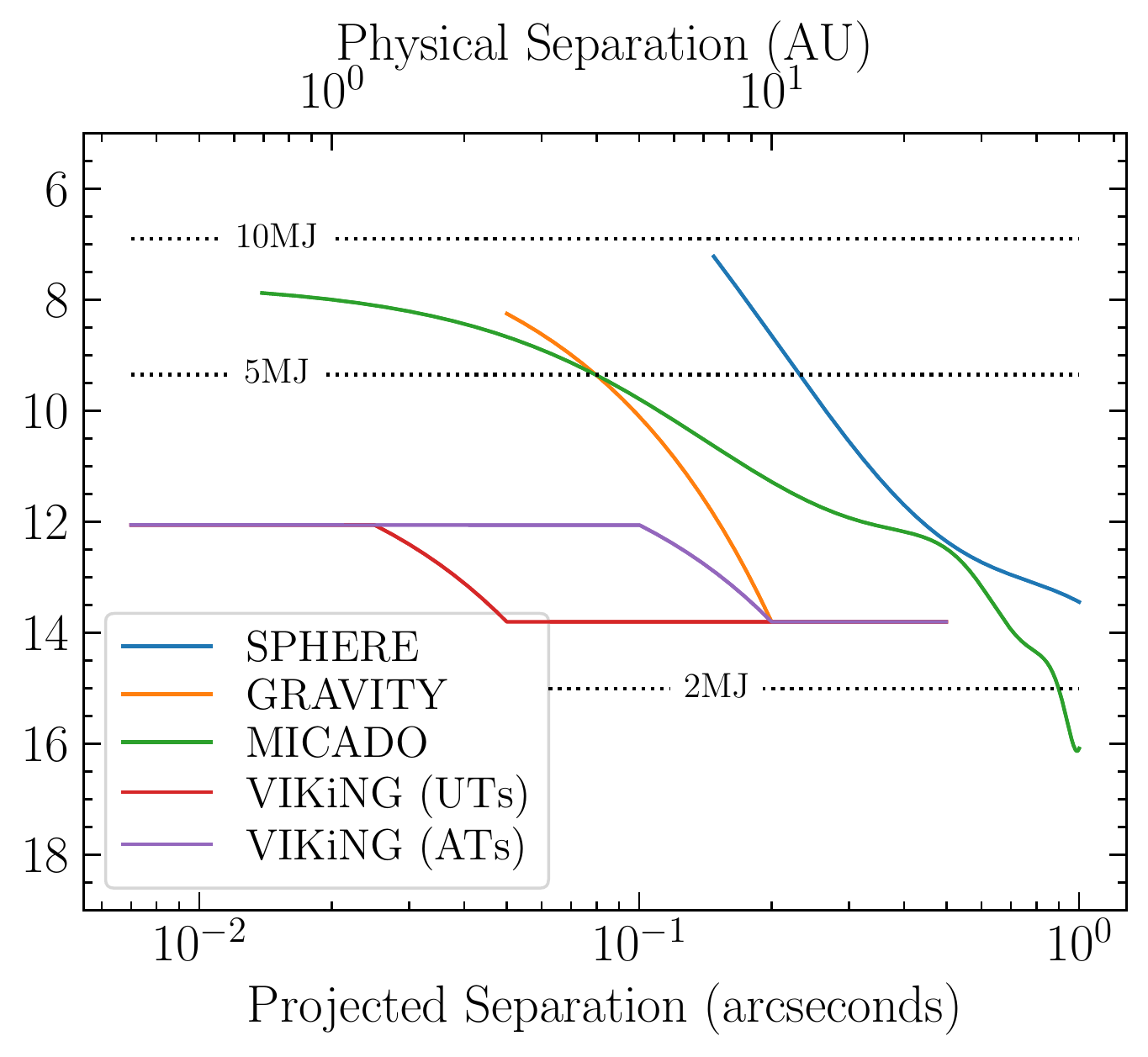}}
    \subfigure[L' band]{\label{fig:l_limits}\includegraphics[width=0.9\linewidth]{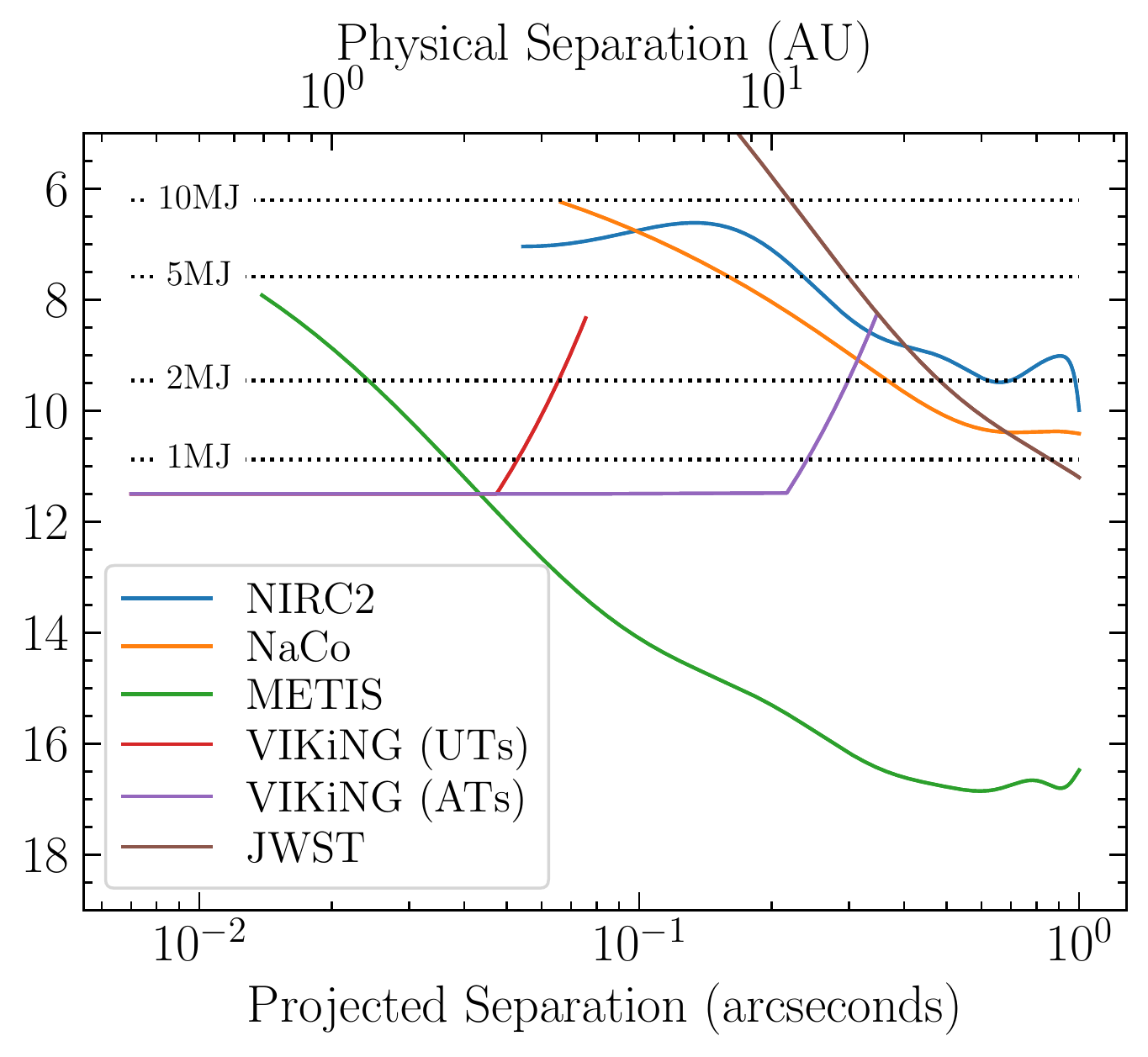}}
    \caption{Assumed limits for current and future instruments as a function of angular and physical separation (assuming an average distance of 50\,pc.) See text for detailed assumptions.  Contrasts of planets of various masses are shown for comparison, assuming an age of 50\,Myr, initial entropy of 10\,$k_{B}$/baryon and stellar magnitude of 7.}
    \label{fig:real_limits}
\end{figure}

The experimental and theoretical contrast limits for current and future instruments are shown in Figure~\ref{fig:real_limits}, assuming a stellar apparent magnitude of 7 in the K and L' bands, which is close to the average magnitude of our targets, and an average distance of 50\,pc.  These limits were determined by observations or, in the case of future instruments, simulated performance. The SPHERE limit is derived from the recent SHINE survey \citep{langlois2021sphere} and the MICADO limits come from \citet{perrot2018design}.  The GRAVITY limit is based on the lower limits of the  curves in \citet{abuter2019hunting}, assuming a 1 hour integration time. The NaCo limits come from the study presented in \citet{quanz2012searching} and the METIS limits are based on \citet{carlomagno2020metis}.  The NIRC2 limit comes from our contrast limit through recent observations of Taurus \citep{wallace2020high}, which are consistent with vortex coronagraph reference star differential imaging limits \citep{xuan18vortex}.  The JWST limit is adapted from \citet{carter2021direct}. 

The VIKiNG limits for L' come directly from the model assuming 1 hour integrations by  \citet{martinache2018kernel}, assuming that the companions must be resolved in the kernel null maps, and with a loss in contrast as the planets approach the edge of the telescope PSF (at separations of 0.5-0.8$\lambda$/D). The assumed contrast of $4\times 10^{-5}$ at 5-$\sigma$ assumes either 80\,nm RMS fringe tracking errors and 10\% intensity fluctuations, or 120\,nm RMS fringe tracking errors and 2\% intensity fluctuations. Note that the Hi-5 instrument \citep{Defrere18} operating in L' is compatible with the VIKiNG architecture, but may require longer integration times for the assumed contrast, depending on the finally adopted architecture. The VIKiNG limits for K are based on a more optimistic, but theoretically possible set of assumptions. For observations with the UTs, fringe tracking up to 0.5" off-axis is assumed with an RMS fringe tracking RMS error of 30\,nm. The contrast limit shown is the magnitude difference between the faintest detectable planet and its star in the K and L' bands.
Figure~\ref{fig:real_limits} demonstrates that interferometry with ViKING and GRAVITY as well as instruments on large telescopes, such as MICADO and METIS are the most sensitive at small ($<$10\,AU) separations.  According to the distribution shown in Figure~\ref{fig:planet_dist}, this is where planets are most likely to occur.  The apparent background-limited magnitude limits of these instruments are shown in Table~\ref{tab:limits}, which apply mostly to faint targets.  We consider a planet to be detectable if it is brighter than both the apparent magnitude limit and has higher contrast than the limit shown in Figure~\ref{fig:real_limits}.
\begin{table}
\caption{Apparent magnitude limits of instruments considered in this paper. Note that for most targets, the contrast and not these background limits most influences detectability.}
\begin{tabular}{lcc}
Instrument & K Mag. Limit & L’ Mag. Limit\\
\hline
SPHERE & 22.5 & - \\
GRAVITY & 19.5 & - \\
MICADO & 29.5 & - \\%Assuming 5 hr integration (Doute 2019)
VIKiNG (UTs) & 23.2 & 18.5\\
VIKiNG (ATs) & 19.9 & 15.3\\ %Come from Martinache & Ireland 2018
NIRC2 & - & 17.94\\ %Comes from 1hr integration with LGS assuming a Strehl ratio of 0.25
NaCo & - & 18.55\\ %Comes from wide angle using exposure time calculator
METIS & - & 21.8\\
JWST & - & 23.8\\ %Comes from 2.5" separation using exposure time calculator
\end{tabular}
\label{tab:limits}
\end{table}

\section{Simulated Detectable Planets}
\label{sec:simulation}
For each star in our sample shown in Section~\ref{sec:star_samp}, we simulate a set of planet systems using the power-law distribution shown in Section~\ref{sec:planet_dist}.  The normalization constant $C$ is scaled linearly with stellar mass.  This simulation was run 5000~times per star, but only a small percentage of these simulations produced planets.  This is due to the integration of the planet distribution shown in Figure~\ref{fig:planet_dist}.  For a 1\,M$_{\odot}$ star, only 8\% of simulations produced a planet more massive than 0.3\,M$_{\mathrm{J}}$.  For simplicity, we assume circular orbits.

\subsection{Planets Detectable by Gaia}
Using our simulated sample of planets around stars in nearby moving groups, we calculated how many can be detected by Gaia applying the methods detailed in Section~\ref{sec:gaia_detect}.  We assumed Gaia can detect planets with periods shorter than 10\,years provided the astrometric signature $\alpha>2\sigma_{\rm{fov}}$ \citep{perryman2014astrometric,ranalli2018astrometry}.  The average number of planets per group and the average number of planets detectable by Gaia are shown in Table~\ref{tab:gaia_detect}, as a result of running the simulation 100 times.

\begin{table*}
\caption{Total number of simulated planets in each group and number detectable by Gaia.  There are $\sim$0.06 planets per star and Gaia can detect approximately $\sim$30\% of these.}
\begin{tabular}{lccccc}
 & & & & Average Number & Average Number\\
Group Name & Age (Myr) & Average Distance (pc) & Number of Stars & of Planets & of Detectable Planets\\
\hline
AB Doradus & 149$^{+51}_{-19}$ & 43.19 & 367 & 19.05 & 6.79\\
Argus & 45$\pm$5 & 48.33 & 630 & 35.10 & 11.32\\
$\beta$ Pictoris & 22$\pm$6 & 39.60 & 149 & 7.82 & 2.91\\
Carina & 45$^{+11}_{-7}$ & 49.22 & 26 & 1.41 & 0.42\\
Carina-Near & $\sim$200 & 34.28 & 148 & 7.50 & 3.23\\
Columba & 42$^{+6}_{-4}$ & 47.47 & 79 & 4.84 & 1.43\\
Hyades & 750$\pm$100 & 46.67 & 239 & 14.97 & 4.47\\
Tucana-Horologium & 45$\pm$4 & 49.34 & 94 & 5.31 & 1.58\\
TW Hydrae & 10$\pm$3 & 53.68 & 21 & 1.00 & 0.32\\
Ursa Major & $\sim$414 & 25.30 & 7 & 0.64 & 0.25\\
\end{tabular}
\label{tab:gaia_detect}
\end{table*}

We have 1760~stars in our sample across all groups shown in Table~\ref{tab:gaia_detect} and, on average, we simulate $\sim$\,98~giant planets ($M>0.3$\,M$_{\mathrm{J}}$) across all groups ($\sim$\,0.056~planets per star.) Our estimates suggest that Gaia is able to detect $\sim$\,33~planets, which is approximately a third of our sample.

\subsection{Planets detectable by both Gaia and Direct Imaging}
\begin{figure}
    \centering
    \subfigure[K band]{\includegraphics[width=0.9\linewidth]{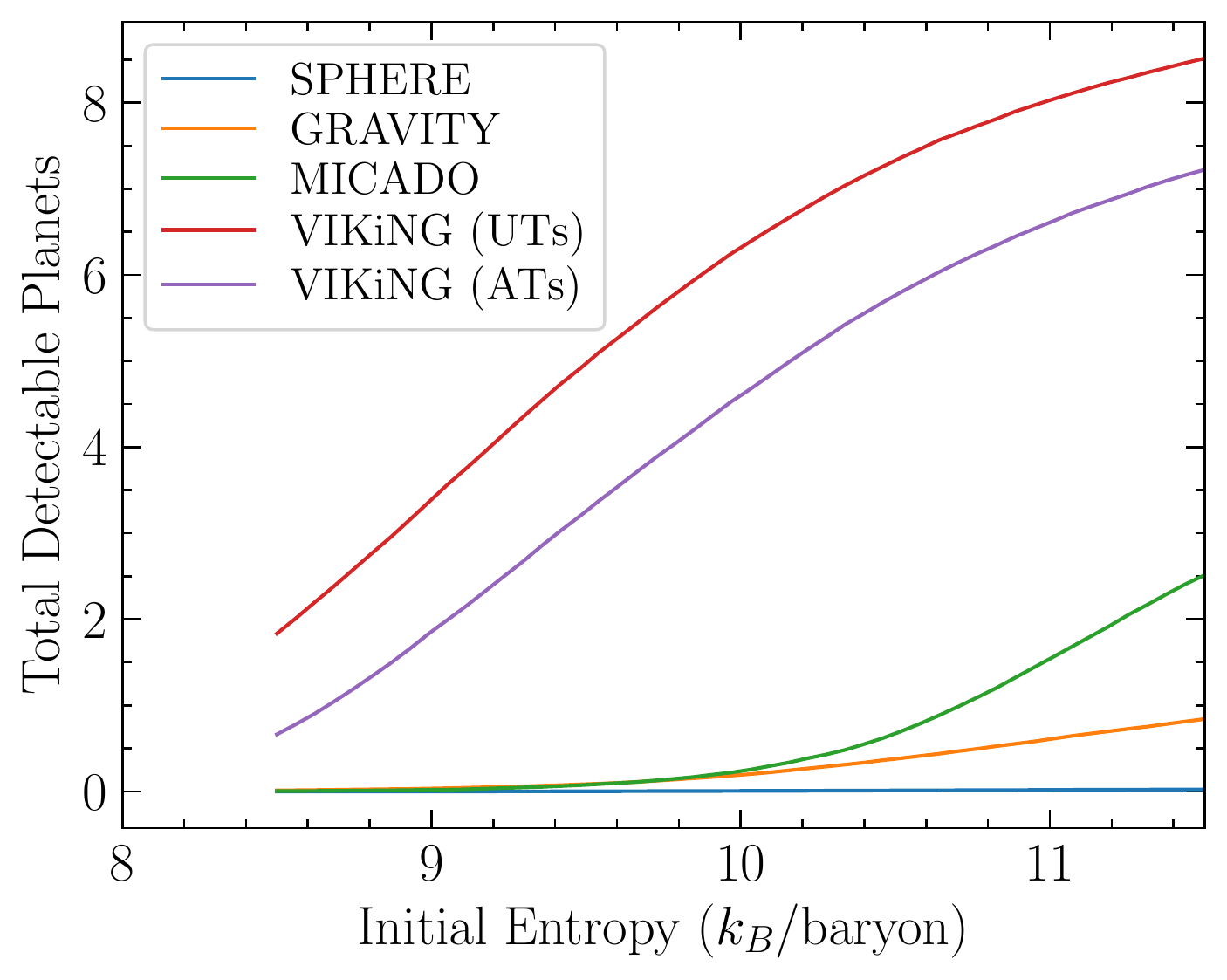}}
    \subfigure[L' band]{\includegraphics[width=0.9\linewidth]{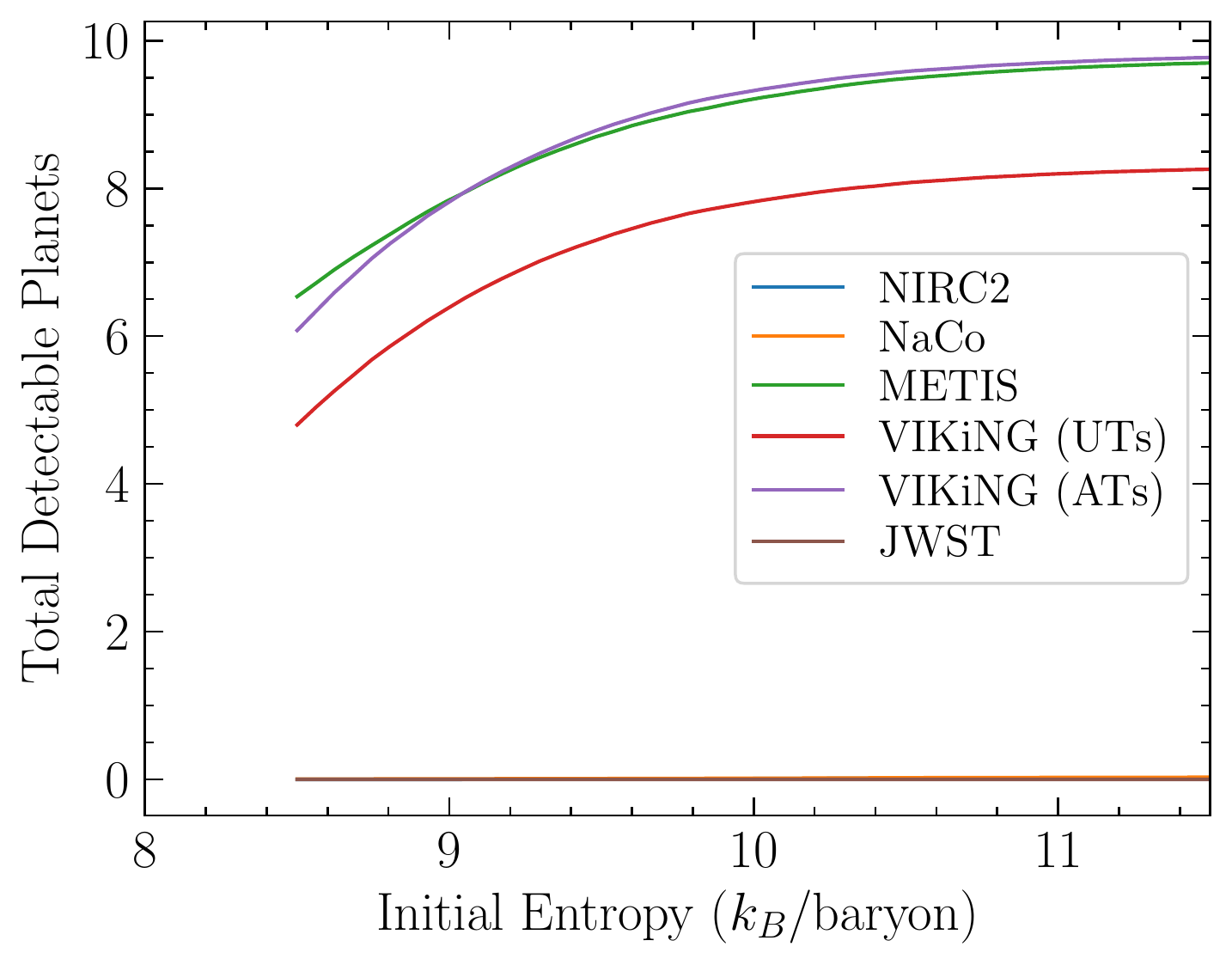}}
    \caption{Total number of planets detectable by both Gaia astrometry and high contrast imaging across all moving groups.}
    \label{fig:detect_planets}
\end{figure}

Using planetary cooling curves shown in Figure~\ref{fig:mag_evol} and the limits of current and future instruments shown in Figure~\ref{fig:real_limits}, we determine how many of the 25 Gaia-detectable planets are also detectable by direct imaging.  This requires an estimate of the age and initial entropy of each planet.  The planet age is assumed to be equal to the age of the moving group, listed in Table~\ref{tab:gaia_detect}, within observational uncertainties and assume a range of initial planet entropies from 8.5--11.5\,$k_{B}$/baryon.  The possible inclination distribution is taken into account by calculating the averaged projected separation.  This is simply calculated by multiplying the simulated semi-major axis by 0.8 as shown in Equation~7 of \citet{fischer1992multiplicity}.  Figure~\ref{fig:detect_planets} shows the number of detectable planets for different instruments in the K and L' bands as a function of initial entropy.
Our results suggest that VIKiNG and METIS should be able to detect more than 4~Gaia-detectable planets regardless of initial entropy, while MICADO should be able to detect hot-start planets, if a survey of nearby moving groups were conducted.  Note that VIKiNG with the ATs can detect more planets than the UTs in the L' band despite being less sensitive.  This is simply due to the wider range of separations detectable by interferometry with the ATs in the L' band as shown in Figure~\ref{fig:l_limits}.

\section{Constraining the Initial Entropy}
\label{sec:constrain}
As explained in Section~\ref{sec:planet_mag}, the initial entropy of a planet is related to its formation conditions and has an effect on the brightness evolution.  If the age and mass of a planet is known to reasonable precision, it should be possible to constrain the initial entropy of a directly imaged planet.

\subsection{Dependence of Magnitude on Entropy}
\begin{figure}
    \centering
    \subfigure[K band]{\includegraphics[width=0.9\linewidth]{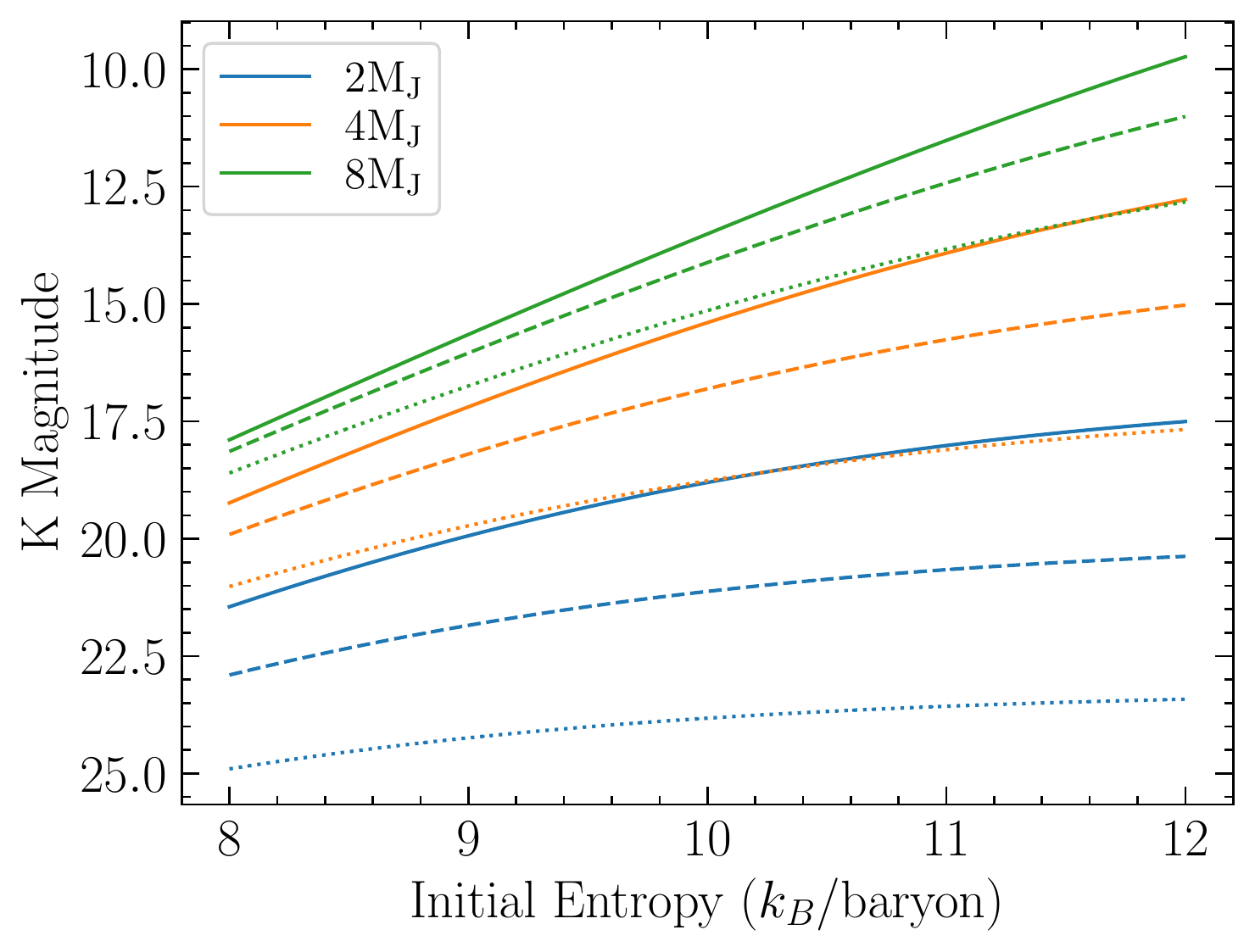}}
    \subfigure[L' band]{\includegraphics[width=0.9\linewidth]{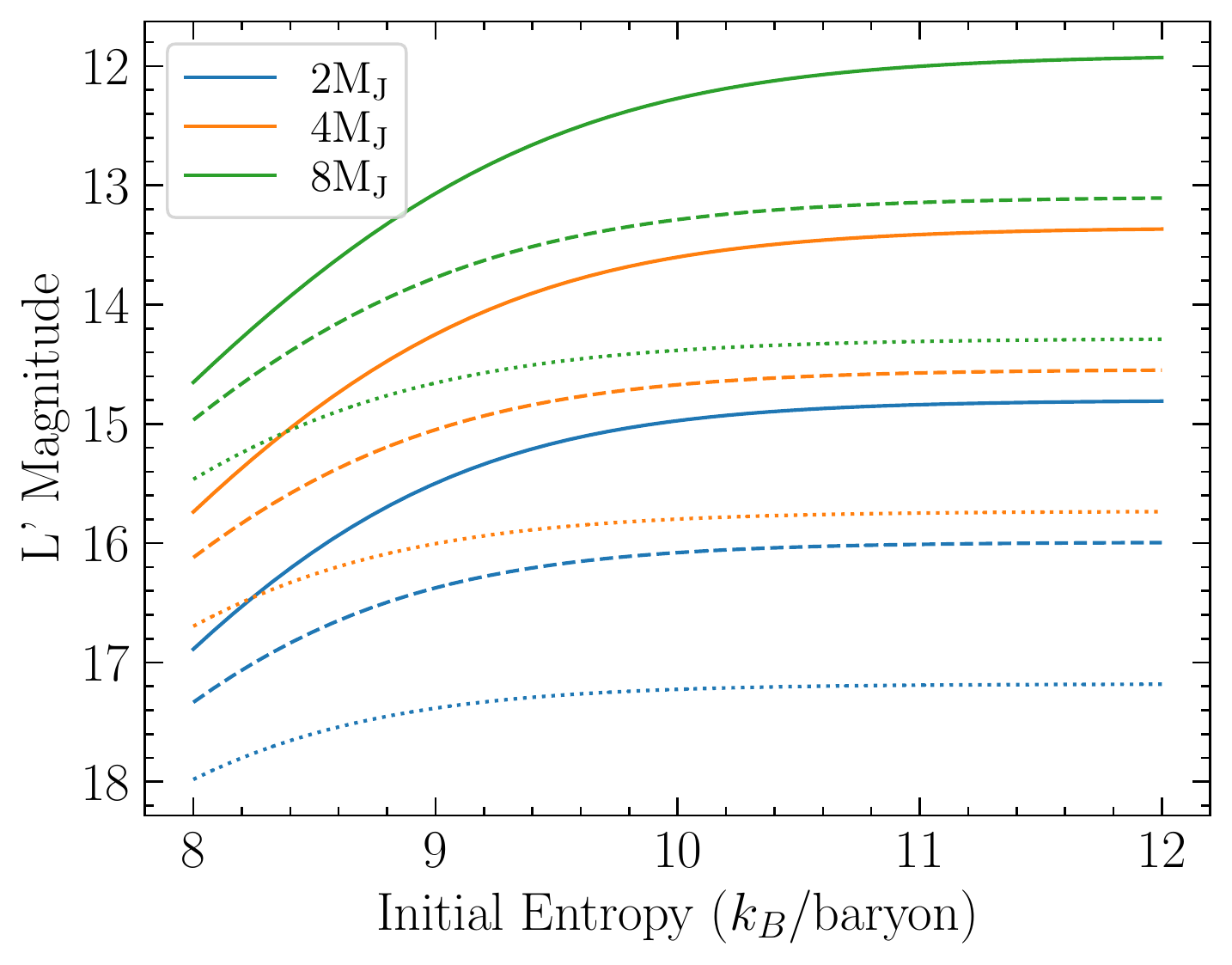}}
    \caption{Absolute Magnitude as a function of initial entropy.  Solid, dashed and dotted curves are for planet ages of 20\,Myr, 40\,Myr and 80\,Myr respectively.}
    \label{fig:mag_v_ent}
\end{figure}

As shown in Figure~\ref{fig:mag_evol}, the magnitude evolution depends on its initial entropy, but this dependence decreases with age.  Figure~\ref{fig:mag_v_ent} shows the absolute magnitude as a function of initial entropy for a variety of planet masses and ages using models from \citet{spiegel2012spectral}.
The curves shown in Figure~\ref{fig:mag_v_ent} flatten as the planets age, but also at higher entropy, particularly in the L' band.  This implies that, while planets with higher initial entropy will be brighter and easier to detect, it could be harder to constrain the initial entropy of these planets.  In order to determine how well different instruments can constrain entropy, we calculate the entropy uncertainty.

\subsection{Entropy Uncertainty}
The entropy uncertainty was calculated for a set of simulated planets detectable by Gaia as shown in Table~\ref{tab:gaia_detect}.  Each planet is assigned a mass and semi-major axis from our distribution, but the initial entropy is unknown.  The likelihood of a particular entropy given our simulated data, $L(S|D)$, is derived from Bayes Theorem,
\begin{equation}
    L(S|D) \propto P(D|S)P(S)\label{eq:likelihood},
\end{equation}
where $P(D|S)$ is the probability of the data for a given entropy and $P(S)$ is the prior probability of that entropy.  The probability as a function of entropy is calculated for a range of entropies $S_{i}$ using a Gaussian distribution in planet flux, given by
\begin{equation}
    P(D|S_{i}) \propto e^{-\frac{(f-f_{i})^{2}}{2\sigma_{f}^{2}}},
\end{equation}
where $f$ is the planet's flux given an input entropy $S$ and $f_{i}$ is the flux for a given entropy $S_{i}$.  The flux error, $\sigma_{f}$, is calculated from the contrast limits of various instruments shown in Figure~\ref{fig:real_limits}.  The 5\,$\sigma$ contrast limits are converted to planet flux limits and divided by 5 to obtain the flux error.  For simplicity, we assume all values of entropy are equally possible and use a flat distribution for $P(S)$.

\begin{figure}
    \centering
    \subfigure[K band]{\label{fig:like_K}\includegraphics[width=0.9\linewidth]{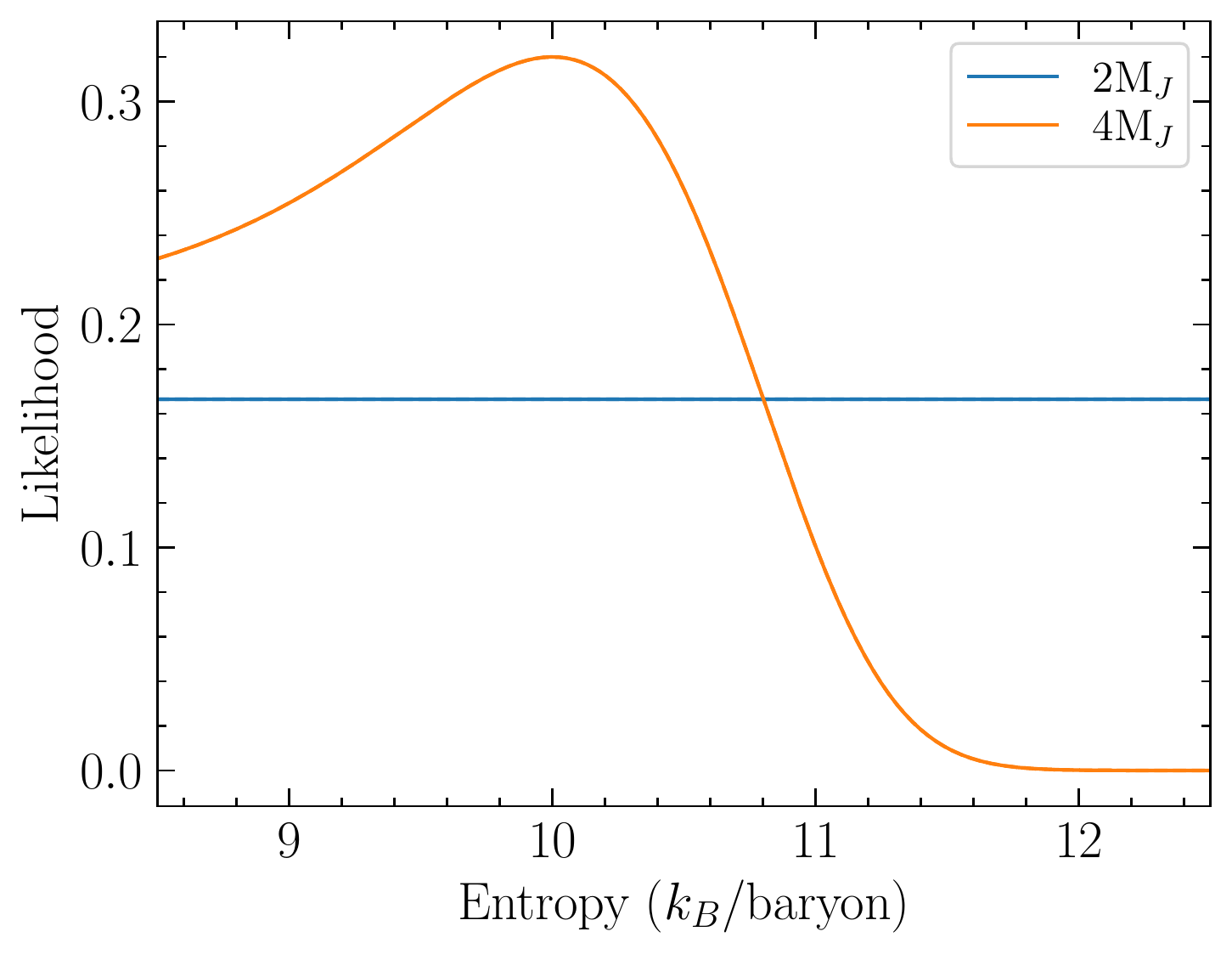}}
    \subfigure[L' band]{\label{fig:like_L}\includegraphics[width=0.9\linewidth]{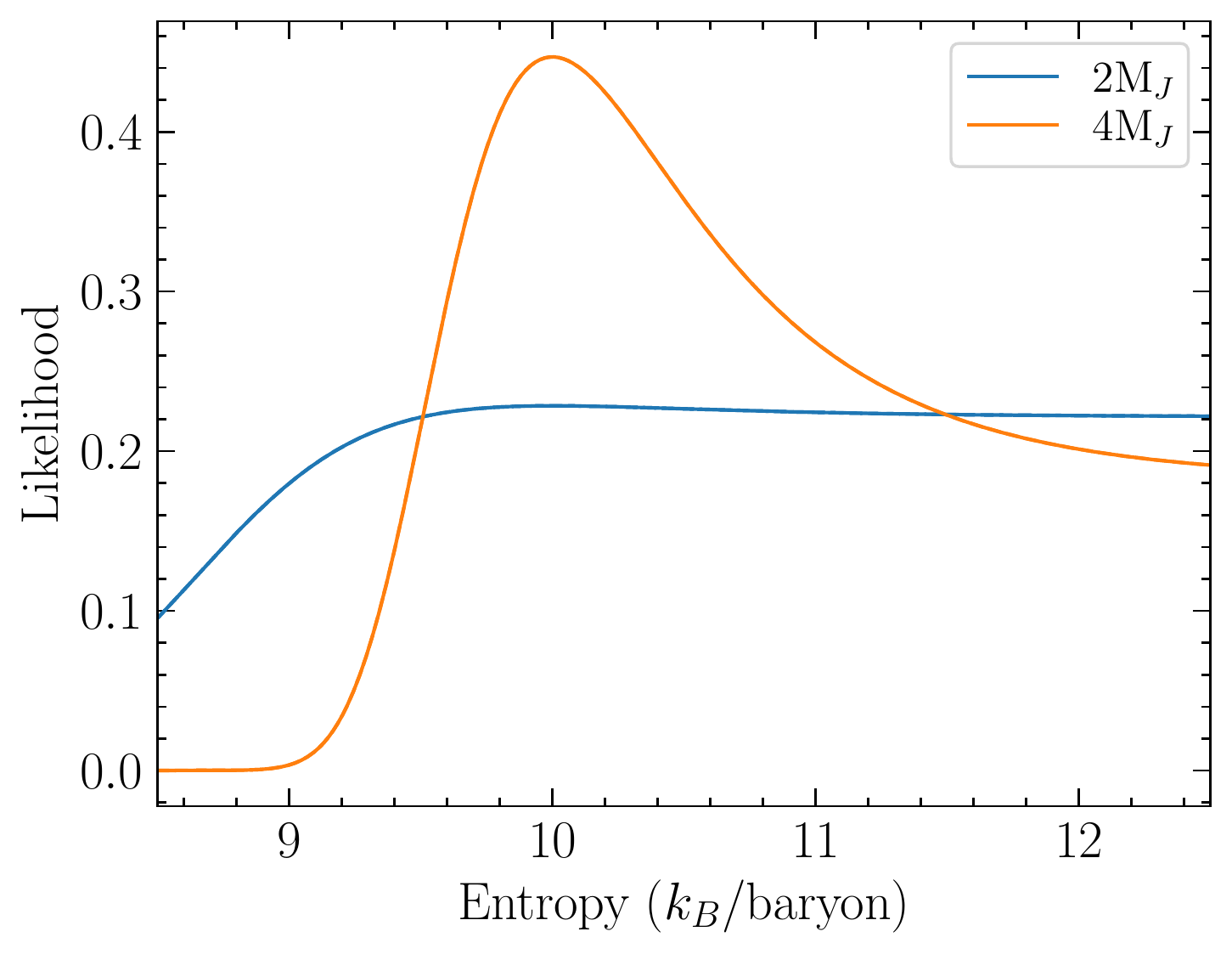}}
    \caption{Likelihood of entropy for given data in K and L' bands, with VIKiNG using the UTs, assuming an input entropy of 10\,$k_{B}$/baryon and a planet 80\,mas from a star of apparent magnitude 7. The solid curves are planets at 10\,au and the dashed curves are planets at 20\,au.}
    \label{fig:likelihood_ex}
\end{figure}

As an example, we consider a star of apparent magnitude~7 in the K and L' bands at a distance of 40\,pc and age of 40\,Myr.  Assuming a true entropy of 10\,$k_{B}$/baryon, the likelihood as a function of modelled entropy is shown in Figure~\ref{fig:likelihood_ex} using VIKiNG with the UTs in both K and L' bands.
The likelihood curves in Figure~\ref{fig:likelihood_ex} confirm that the initial entropy of a 2\,M$_{J}$ planet cannot be constrained, while the entropy of a 4\,M$_{J}$ can be broadly constrained in the L' band, but not in the K band.

\begin{figure*}
    \centering
    \subfigure[K band]{\includegraphics[width=0.4\linewidth]{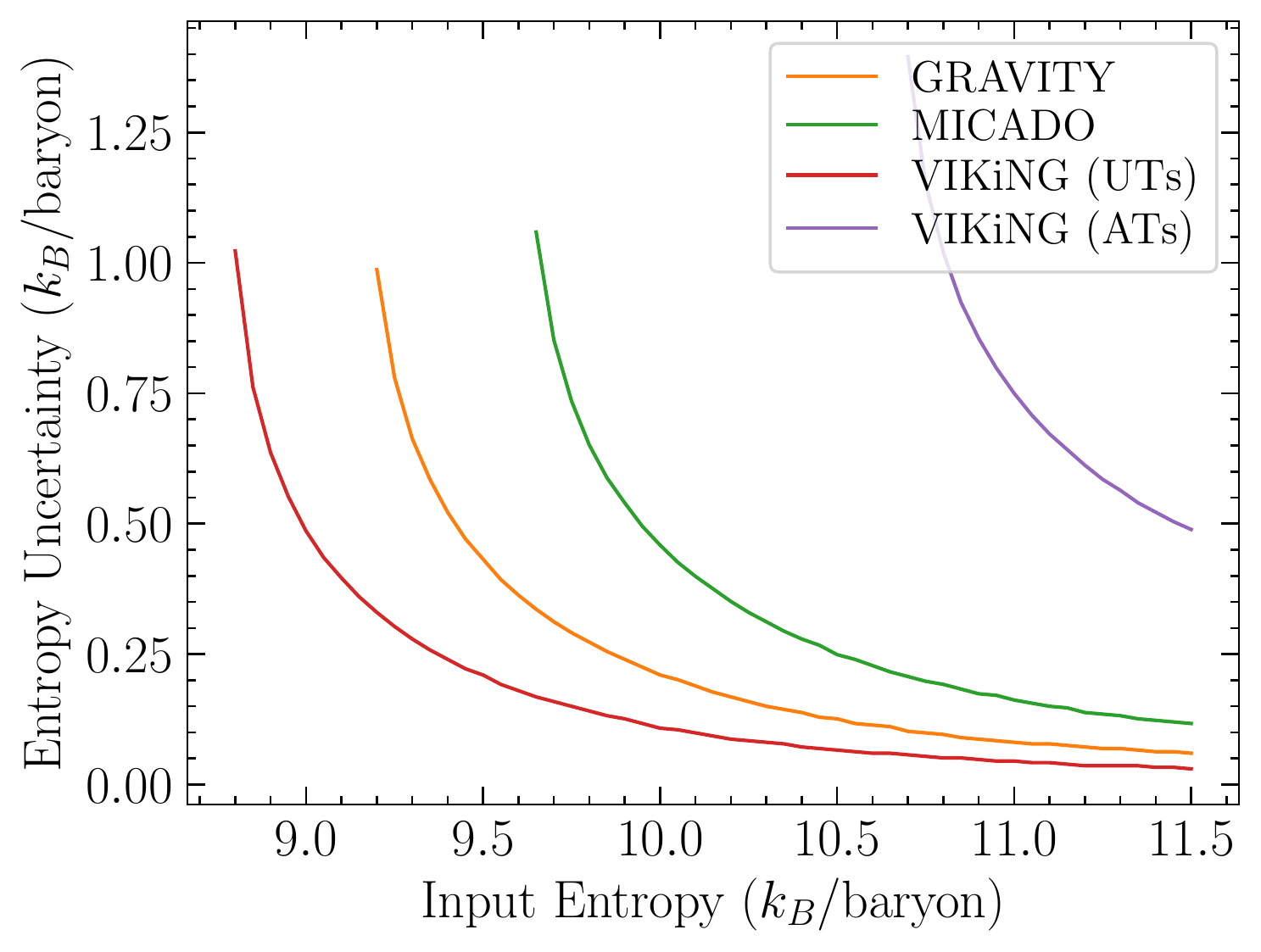}}
    \subfigure[L' band]{\includegraphics[width=0.4\linewidth]{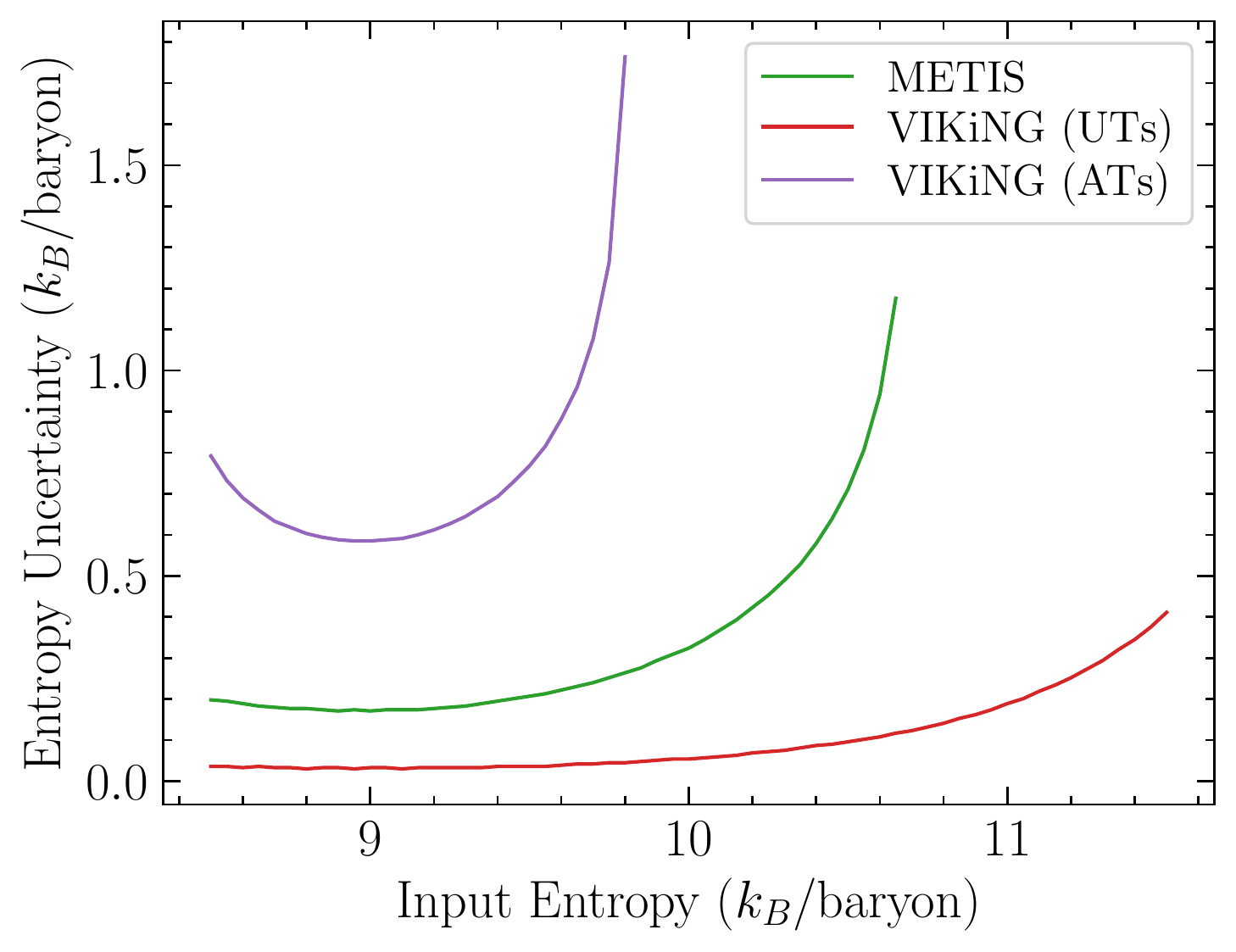}}
    \caption{Entropy Uncertainty as a function of input entropy.  The horizontal axis is a simulated entropy which translates to a theoretical brightness and the vertical axis is the width of the entropy likelihood function as shown in Figure~\ref{fig:likelihood_ex}.  For VIKiNG with the UTs, the entropy uncertainty is below 0.5\,$k_{\rm{B}}$/baryon for all reasonable values of initial entropy which is less than the difference between hot and cold-start models.  This implies VIKiNG should be able to distinguish between the two models.}
    \label{fig:ent_result}
\end{figure*}

This method was applied to a random sample of Gaia detectable planets.  Each planet was assigned a set of initial entropies from 8.5--11.5\,$k_{B}$/baryon and the likelihood function was calculated for each of these.  From this, we calculated the entropy uncertainty as a function of input entropy.  As shown in Figure~\ref{fig:detect_planets}, only GRAVITY, MICADO, METIS and VIKiNG will be able to detect at least 1~planet that is also detectable by Gaia.  The majority of simulated planets have entropies that cannot be constrained.  As a preliminary result, we only consider 1~planet that is detectable by Gaia and all of the instruments mentioned above.  The result for a 3.6\,M$_{J}$ planet 2.4\,AU from a 0.75\,M$_{\odot}$ star in the $\beta$-Pic moving group is shown in Figure~\ref{fig:ent_result}.
The curves in Figure~\ref{fig:ent_result} indicate that VIKiNG with the UTs will be able to constrain the initial entropy within 0.5\,\,$k_{\rm{B}}$/baryon for the majority of input entropies.  GRAVITY and METIS can also constrain the entropy within this value for a `warm'-start planet (entropy of 9--10\,$k_{B}$/baryon), which indicates these instruments should be able to distinguish between hot and cold-start models for this planet.  We note that, unlike the other instruments, VIKiNG is still only at a preliminary design study level at this point.

% =================================================
\section{Conclusions}
\label{sec:conclusion}
In this paper, we have examined the brightness evolution of giant planets and the dependence on initial entropy.  Given that high-entropy planets are brighter than low-entropy planets of similar mass but this difference becomes less at old ages.  If we observe planets at young ages, it should be possible to constrain initial entropy.

Using the expected uncertainty for Gaia astrometry from \citet{perryman2014astrometric}, we determine that Gaia should be able to detect approximately 25\,\% of giant planets in nearby moving groups and calculate their mass.  Combining this with the estimated ages of these moving groups, we can use the detected flux to determine the initial entropy.  We performed this over a simulated sample of planets around existing stars in nearby moving groups, assuming a symmetric planet distribution from \citet{fernandes2019hints}.

We used the measured and expected 5$\sigma$ contrast limits for current and future instruments to estimate the expected flux error on the planets in our simulated sample.  We found that future instruments MICADO and METIS have the best contrast levels at wide angles, while interferometers GRAVITY and VIKiNG are best at small angles.  However, we note that improvements to GRAVITY, known as GRAVITY+ are currently being implemented and, of the VIKiNG concepts, only the L' concept has significant funding. Given the technological challenges of achieving the required 30\,nm fringe tracking uncertainty for a VIKiNG K and that entropy is better constrained by VIKiNG L', this paper does not provide a reason to prioritise a high performance Nuller for VLTI operating in the K filter. Overall, assuming Gaia can detect giant planets in nearby moving groups, we find that these future instruments should also be able to detect some planets, if we were to conduct a survey of a relatively small number of Gaia-detected planets in nearby moving groups.

Using the instrumental flux error, combined with the estimated ages and masses from Gaia, we can constrain the formation entropy of directly imaged planets.  We found that GRAVITY, METIS and VIKiNG should all be able to constrain the formation entropy of a super-Jupiter to within 0.5\,$k_{B}$/baryon and from this, distingush between hot and cold-start formation models.

\section*{Data Availability}
The data underlying this article are available from the corresponding author on reasonable request.

\section*{Acknowledgements}
We thank the anonymous referee for their useful comments, which greatly improved this study and the organization of this paper.
We thank Mark Krumholz for initiating the discussions of planet entropy, which led to this work.  This research was supported by the Australian Government through the Australian Research Council's Discovery Projects funding scheme (DP190101477). C.~F.~acknowledges funding provided by the Australian Research Council through Future Fellowship FT180100495, and the Australia-Germany Joint Research Cooperation Scheme (UA-DAAD).

%%%%%%%%%%%%%%%%%%%%%%%%%%%%%%%%%%%%%%%%%%%%%%%%%%



\renewcommand\refname{References}
\bibliographystyle{mnras}
\bibliography{references} % if your bibtex file is called example.bib

%\begin{thebibliography}{99}
%\end{thebibliography}

%%%%%%%%%%%%%%%%%%%%%%%%%%%%%%%%%%%%%%%%%%%%%%%%%%

%%%%%%%%%%%%%%%%% APPENDICES %%%%%%%%%%%%%%%%%%%%%

%\appendix


%%%%%%%%%%%%%%%%%%%%%%%%%%%%%%%%%%%%%%%%%%%%%%%%%%

% Don't change these lines
\bsp	% typesetting comment
\label{lastpage}
\end{document}